\documentstyle[11pt,epsfig]{article}
\advance\textheight by \topskip
\begin{document}
\vspace*{\fill}
\begin{center}
{\Large \bf 
Gluino signals in 4jet events\\
and vertex tagging at LEP~I}\\[1.5cm]
{\large Stefano Moretti$^{a,b}$, Ramon Mu\~noz-Tapia$^c$,}\\[0.25 cm]
{\large and Kosuke Odagiri$^{b}$}\\[0.4 cm]
{\it a) Dipartimento di Fisica Teorica, Universit\`a di Torino,}\\
{\it and I.N.F.N., Sezione di Torino,}\\
{\it Via Pietro Giuria 1, 10125 Torino, Italy.}\\[0.25cm]
{\it b) Cavendish Laboratory, University of Cambridge,}\\
{\it Madingley Road, Cambridge CB3 0HE, UK.}\\[0.25cm]
{\it c) Dpto F\'{\i}sica Te\'orica y del Cosmos,}\\
{\it Universidad de Granada, Granada 18071, Spain}\\[0.5cm]
\end{center}
\vspace*{\fill}

\begin{abstract}
{\noindent 
\small 
Heavy flavour tagging provides a broad range of possibilities
in testing QCD features at LEP. We present here a 
study of 4jets events at LEP I where the so-called light
gluinos could be directly produced.  We show that microvertex techniques
offer a unique chance to exploit simple kinematical distributions in order
to optimise the signal coming from gluino production with respect
to the background of ordinary QCD events. 
Our results indicate that experimental analyses along the lines
suggested here can exclude or reveal the presence of a 
gluino for masses up to 10 GeV and lifetimes below 
10$^{-9}$ sec. We also point out that a large fraction of gluino
events could decay in configurations carrying large missing energy, so to
escape the usual selection criteria of 4jet samples.
In our study, mass effects of quarks and gluinos have been taken into
account exactly. Our results are independent from both the 
jet algorithm and its resolution parameter.}
\end{abstract}
\vskip1.0cm
\hrule
\vskip0.25cm
\noindent
Electronic mails: moretti@hep.phy.cam.ac.uk, rmt@ugr.es,
                  odagiri@hep.phy.cam.ac.uk.

\vspace*{\fill}
\newpage

\section{Introduction}

Supersymmetry (SUSY)  predicts the existence of a spin 1/2 partner
of the gluon $g$, the so-called { gluino} $\mathaccent"7E g$. 
Like the QCD gauge boson, it is neutral, it is its own anti-particle (i.e.,
it is a Majorana fermion) and its coupling to ordinary matter is precisely
determined in terms of the usual QCD colour matrices and the strong
coupling constant $\alpha_s$ \cite{Nilles}. 
However, in contrast to the gluon, whose mass is predicted to be zero 
by the theory, that of the gluino $m_{\mathaccent"7E g}$
 is a priori an arbitrary 
parameter, and so is its lifetime $\tau_{\mathaccent"7E g}$.

Many searches have been carried out in order to detect or rule out
such a particle. A detailed survey, including the description
of various experiments, can be found in 
Ref.~\cite{Farrar-review}. In particular, light gluinos should be
directly produced in 4jet events at LEP~I.

Motivated by the advances in 4jet analyses based on heavy flavour
identification, we further elaborate the study of  Ref.~\cite{physlett}
to cater for a wider range of light gluino masses and lifetimes.
The use of $\mu$-vertex devices \cite{deangelis,Squarcia} provides an
independent procedure to
settle the ongoing controversy around the light gluino scenario,
if one considers that long-lived gluinos might produce 4jet events with
detectable secondary vertices \cite{vertex,Cuypers}. We assume them to 
be long-lived enough to be tagged by present experimental vertex tagging
methods.
With simple invariant mass cuts based on the different kinematics of
the partons in the final state, one can obtain a signal identifiable as
a clear excess in the total number of vertex 
tagged 4jet events, in percentage well
beyond the uncertainties related to non-perturbative 
as well as to higher order perturbative effects.
We  also consider the possibility
of `loopholes' in recent analysis carried out by the ALEPH 
collaboration~\cite{aleph} and by de Gouv\^ea and Murayama~\cite{Murayama},
which could undermine the validity
of the results obtained there.
We are especially concerned with the fact that gluinos might decay mainly 
into missing energy, so that the $q \bar q \tilde g \tilde g$ events are not
recognised as 4jet events. As a matter of fact, the ALEPH analysis explicitly 
assumes that this is not the case whereas Ref.~\cite{Murayama} only considers
the case in which the gluino does not decay inside the LEP detectors.

The plan of the paper is as follows. In next Section we 
review the status of the light gluino window and put our work into context. 
Next we describe
our calculations, and in Section 4 we devote some space to discuss possible
tagging procedures of SUSY events.
In Section 5 we present our results and Section 6 contains some brief 
conclusions.

\section{The light gluino window}

Our present knowledge about gluinos is summarised in 
Fig.~1, which shows the excluded regions in mass and lifetime of the SUSY 
fermion 
as they stood in 1993-1994\footnote{We 
would like to thank the authors of
Ref.~\cite{Kileng-Osland1} for their kind permission of exploiting 
here one of the figures of their paper.}.
At that time, it was clear that relatively long-lived and light gluinos
(decaying into a `photino', or more correctly, into the
lightest neutralino)
were not yet excluded by the experiment.

In the theory it is natural for gluinos to be much lighter than squarks
if their mass is induced radiatively \cite{Masiero}. Furthermore, 
gluino and lightest neutralino masses are naturally
less than few GeV if dimension-3 SUSY-breaking operators are
absent from the low energy theory \cite{Farrar-Quarks94}, although
the compatibility with a light gluino window in 
Supergravity models is strongly dependent on the dynamics 
of the breaking mechanism
of the electroweak symmetry \cite{Diaz,Burjassot}.
Since these values of $m_{\mathaccent"7E g}$ 
and $\tau_{\mathaccent"7E g}$ were within the reach
of already operating accelerators \cite{PDG}, the regions identified 
by the white areas in Fig.~1 started receiving 
some attention in those years \cite{window}.
In particular, it was noted that if the gluino is so light, it should be 
directly produced at LEP I: either in 2jet \cite{Kileng-Osland2} or
in 4jet events \cite{epem}.

An immediate interest in this possibility raised. This was also
motivated by the `historical' 
discrepancy between the value of $\alpha_s$ determined by
low energy deep-inelastic lepton-nucleon scattering and that measured
by the $e^+e^-$ CERN experiments \cite{alphas}.
In this respect, although the discrepancy
between the two values of $\alpha_s$ was statistically small
\cite{cpl} and also has slightly decreased in the latest 
measurements \cite{last_as},
it was speculated that the evolution of the strong
coupling can be slowed down by a contribution to the $\beta$ function
of a new, coloured, neutral fermion: indeed, a light gluino.

The search for SUSY-signals intensified then at LEP I. 
Studies in other contexts, such as the influence in the Altarelli-Parisi
evolution of the structure functions \cite{GLAP} or the so-called
`3+1' jet events at HERA \cite{Ramon}, were also pursued. However, the
effects are there too small to be tested using 
the present experimental data. More recently,  
extensive searches for light
gluino signals have been carried out at the Tevatron \cite{CMF}.
As for LEP I,
the strategy adopted was to search for light gluinos 
in the context of the so-called QCD colour factor analyses in 4jet 
samples \cite{colour}. 
The basic idea is to measure the fundamental colour factors of QCD, that is,
$C_A$, $C_F$ (the Casimir operators of the fundamental and adjoint 
representations of
the gauge group $SU(N_C)$) and $T_F$ (the normalisation of the
generators of the fundamental representation). 
In QCD (i.e., $N_C=3$), one gets $C_A=3$ and $C_F=4/3$.
The factors $C_A$, $C_F$ and $T_F$ represent the relative
strength of the couplings of the processes $q\rightarrow qg$, $g\rightarrow gg$
and $g\rightarrow q\bar q$, respectively (see, e.g., Ref.~\cite{Quigg}).
The analytical formulae 
of the cross section $\sigma(e^+e^-\rightarrow 4\mbox{jet})$ 
for massless particles 
were computed long ago in Ref.~\cite{ERT}. 
The strategy is to compare 
the theoretical predictions to the data, by 
leaving the colour factors as free parameters to be determined by
the fit. 
In practice, one of these, e.g., $C_F$, is absorbed in the normalisation
of the cross sections leaving two independent ratios $C_A/C_F$ and 
$T_R/C_F$,
with $T_R=N_FT_F$,
being $N_F$ the number of active flavours.


The experimental analyses are preferentially based on angular correlations 
between jets \cite{angles}, as they are sensitive 
to differences between the $2q2g$ (Fig.~2a--c) and the 
$4q$ (Fig.~2d) component of 4jet events. 
Light gluinos would enter in 4jet events via diagrams of
the type depicted in Fig.~3, in the process $e^+e^-\rightarrow q\bar q
{\mathaccent"7E g}{\mathaccent"7E g}$ (through a $g^*\rightarrow
{\mathaccent"7E g}{\mathaccent"7E g}$ 
splitting). Note that gluino production in 4jet events via squarks splitting into
quark-gluino pairs is very  suppressed due to 
the large value of the lower limits on the squark masses, 
so is also the case  in 2jet production through squark loops 
\cite{Kileng-Osland2}.
As gluinos are coloured fermions, their contribution would 
enhance the part of the 
4jet cross section with angular structure similar to that of
$4q$ events.
Naively then, one could say that the
total number of flavours $N_F$ of the theory is apparently
increased, such that, a
SUSY-signal can be revealed in the form of an enhancement of 
$T_R$, with respect to the predictions of pure QCD.
The results of those analyses were that, 
although the experimental measurements were in good agreement with QCD,
it is was not possible 
to exclude the existence of a light ${\mathaccent"7E g}$ (see, e.g., 
Ref.~\cite{OPAL}). In particular,
gluinos with a mass of at least
2 GeV yield an expectation value for $T_F/C_F$ 
that was within one standard deviation of the measured one. Even the 
extreme case of a massless gluino (for which $T_F/C_F\approx0.6$) would have
brought the predictions only slightly beyond the upper experimental region
of $68\%$ confidence level (CL) given in Ref.~\cite{OPAL}. Therefore,
after those studies, the experimental constraints on the gluino mass
and lifetime could still be summarised by the plot in Fig.~1.

The reason why the LEP analyses showed a
limitation in putting stringent bounds on the existence of light
gluinos was that contributions to the total cross section
of SUSY events are small and further reduced with respect to the 
ordinary QCD rates 
when mass suppression is taken into account. In particular, gluino effects on 
the total number of 4jet events were comparable in percentage to the 
systematic uncertainties related to jet hadronisation process and 
the uncalculated 
next-to-leading order (NLO) corrections through the order 
${\cal O}(\alpha_s^3)$.

To overcome these systematic limitations, it was recently proposed in 
Ref.~\cite{Murayama} to consider the colour factor
$C_A/C_F$ sufficiently well known as to be taken for its QCD value, 9/4.
The other factor is also partially constrained, by the fact that there are
clearly five quark flavours which are active in the di-jet $Z$ decays.
Therefore, if one wants to pin down possible gluino effects, one should
allow variations of $T_F/C_F$ only above the value 3/8. Armed with these two 
new constraints, Ref.~\cite{Murayama} obtained an improved bound on 
$m_{\mathaccent"7E g}$. In particular, it was shown that a gluino mass  
${\buildrel{\scriptscriptstyle <}\over{\scriptscriptstyle\sim}}~1.5$ 
GeV is apparently excluded at more than $90\%$ CL 
by the 1991-92 OPAL data \cite{OPAL}.
A new experimental analysis carried out by ALEPH
in Ref.~\cite{aleph} obtains results along the same lines.
They give new measurements of the QCD colour factors using
all the data collected from 1992 to 1995 and
obtain excellent agreement with ordinary
QCD along with a new 95\% CL constraint $m_{\mathaccent"7E g}>6.3$ GeV 
on the gluino mass\footnote{Further indications towards the exclusion
of somewhat lighter gluino masses come from studies of the running
of $\alpha_s$ at higher orders \cite{as3}.}.
This  stringent limit was achieved
thanks to a dedicated treatment to reduce 
Monte Carlo (MC) uncertainties related
to the hadronisation process of the partons and to the fact that meanwhile
preliminary results of the NLO corrections to the 4jet rate had become
available (the complete calculation
has been presented very recently \cite{a3complete}).
On the one hand, several different models of parton
fragmentation were compared to each other with different parameter settings,
on the other hand, it was clear that NLO results 
have a strong impact on the 4jet rate, but very
small influence on the shape of the angular distributions used
in Ref.~\cite{aleph}.

Although these results represent a clear improvement, the 
analyses of Refs.~\cite{aleph,Murayama}
are still based on the traditional method \cite{colour} of 
ordering the jets in 
energy and identifying the two most energetic ones with those originated
by the quarks produced in the  
$Z$ decay. In fact, the angles which are generally used
(see Ref.~\cite{angles} for the exact definitions) 
in measuring the colour factors of QCD  require in principle
the identification of primary and secondary partons. In practice, the 
above assumption is often incorrect and the sensitivity of the 
experimental distributions to the QCD  colour factors is considerably reduced. 
In this respect, it is worth recalling that, e.g., in 
$Z\rightarrow q\bar qgg$ events the percentage 
of events in which the two lowest energy partons are both gluons is 
only $\approx53\%$ \cite{nonAbel}. 

A possible improvement of the `energy ordering' procedure was advocated in 
Ref.~\cite{ioebas}, where samples of 4jets with two jets tagged as heavy
flavour jets (i.e., $c$- 
and especially $b$-quarks) \cite{deangelis} are considered.
In this way, one gets a greater 
discrimination power between $q\bar qgg$ and $q\bar q
q'\bar q'$ events, for two reasons.
First, one is able to distinguish
between (heavy) quark and gluon jets,  thus assigning the momenta of the final
states to the various particles in a more correct way, as heavy quarks are
mainly produced as primary partons.
Second, $q\bar qgg$ event rates are reduced by the heavy flavour selection
by a factor
of 3/5 with respect to the $q\bar q q'\bar q'$ ones.
Therefore, the $4q$ signal is enhanced by almost a factor of two 
and the differences between the quark and the gluon components 
can be more easily studied. 

The DELPHI collaboration is the only one to date (to our knowledge) that 
has resorted to flavour identification techniques to
analyse 4jet events \cite{ana} (similar studies in the
case of 3jet events are performed in  Ref.~\cite{fuster}). 
They examined the data collected in the years 1991-1994 , from
which a total of 11,000 4jet events with at least two heavy quark jets were 
selected. The typical efficiency was 12\% with a purity
of events where all jets are correctly assigned of about 70\%. Note 
that  neural networks were employed to combine the 
information on high transverse momentum leptons, large impact parameters
and energy ordering of the jets (see, e.g., Ref.~\cite{deangelis}
for a review about techniques of heavy flavour identification).
Their results have been presented recently 
\cite{done}. The important outcome is that with the new selection strategy 
the errors are substantially reduced compared to previous analyses
\cite{colour,OPAL}, especially for $T_R/C_F$.
In general, the result was found
to be in good agreement with the QCD expectations, but no new constraint
on the mass of a possible light gluino was given at that time. Further
analyses along the same lines are currently in progress \cite{progress}.
As we shall see later on, our findings further  support the relevance 
of such approaches.

\section{Calculation} 

The Feynman diagrams describing at tree-level the reactions
\begin{equation}\label{proc1}
e^+ + e^-\rightarrow q + \bar q + g + g,
\end{equation}
\begin{equation}\label{proc2}
e^+ + e^-\rightarrow q + \bar q + {q'} + \bar q',
\end{equation}
\begin{equation}\label{proc3}
e^+ + e^-\rightarrow q + \bar q + {\mathaccent"7E g} + {\mathaccent"7E g},
\end{equation}
are shown in Figs.~2--3. In the present analysis
we have computed the matrix elements of processes (\ref{proc1})--(\ref{proc3})
with the same {\tt FORTRAN} generator used in
Refs.~\cite{ioebas,MEs}, which takes exactly into account all masses
and both the $\gamma^*$ and $Z$ intermediate contributions. Mass effects 
in 4jet events are important, as repeatedly  recalled in the 
literature \cite{ioebas,MEs,masses,Mainz}, especially if heavy flavour selection
is performed.
For this study, the above program has been also checked
against the one used in Ref.~\cite{masses} in the appropriate
limit (i.e., when the masses along the fermion lines attached to 
the $\gamma^*,Z$ propagator are neglected).
For the details of the numerical computation
as well as the explicit helicity amplitude formulae, see Ref.
\cite{MEs}. 

We have analysed processes (\ref{proc1})--(\ref{proc3}) adopting
four different jet-resolution criteria for resolvable partons. 
We have done so in order to 
investigate the independence of our conclusions
from the actual criteria employed and to check whether any of these
shows better features for the
analysis of light gluino contributions.
The jet algorithms are identified through
their clustering variable $y_{ij}$. They are ($\sqrt s=M_Z$): 
the JADE scheme (J) \cite{JADE}
based on the `measure'
\begin{equation}\label{JADE}
y^J_{ij} = {{2E_i E_j(1-\cos\theta_{ij})}\over{s}},
\end{equation}
and its `E' variation (E)
\begin{equation}\label{E}
y^E_{ij} = \frac{(p_i+p_j)\cdot(p_i+p_j)}{s},
\end{equation}
the Durham scheme (D) \cite{DURHAM}
\begin{equation}\label{DURHAM}
y^D_{ij} = {{2\min (E^2_i, E^2_j)(1-\cos\theta_{ij})}
\over{s}}
\end{equation}
and the Geneva algorithm (G) \cite{GENEVA}
\begin{equation}\label{GENEVA}
y^G_{ij} = \frac{8}{9} {{E_iE_j(1-\cos\theta_{ij})}\over{(E_i+E_j)^2}}.
\end{equation}
For all of them the two (pseudo)particles $i$ and $j$
(with energy $E_i$ and $E_j$, respectively) for which $y_{ij}$
is minimum are combined into a single (pseudo)particle $k$ of momentum
$P_k$ given by the formula
\begin{equation}
P_k=P_i+P_j.
\end{equation}
The procedure is iterated until all (pseudo)particle pairs satisfy $y_{ij}\ge
y_{\mathrm{cut}}$. The various characteristics of these algorithms are 
summarised in Ref.~\cite{GENEVA}. In our lowest order calculation,
the 4jet cross section for a given algorithm is simply equal to
the four parton cross section with a cut $y_{ij}\ge y_{\mathrm{cut}}$ on all
pairs of partons $(i,j)$.

It is worth noticing that, in the recently calculated NLO  corrections
to the 4jet rates \cite{a3complete}, the much forgotten Geneva scheme 
has been shown to be particularly sensitive to the number of light flavours
as well as to exhibit a small scale dependence. In this respect, the 
G scheme may be more suitable than others in enlightening possible 
gluino contributions in the experimental sample.
 
Concerning the numerical part of our work, we have taken
$\alpha_{em}= 1/128$ and  $\sin^2\theta_W=0.23$, while
for the $Z$ boson mass and width we have adopted the values
$M_{Z}=91.1$ GeV and $\Gamma_{Z}=2.5$ GeV, respectively.
For the quarks we have: $m_c=1.7$ GeV and $m_b=5.0$ GeV while
the flavours $u$, $d$ and $s$ have been considered massless.
We have varied the gluino mass $m_{\mathaccent"7E g}$ 
in the range between 0 and
20 GeV. Finally, the strong coupling constant has been 
set equal to $0.115$\footnote{Our results will not be 
affected by the actual value of $\alpha_s$, as
we will be interested in the end in relative differences between
ordinary QCD and QCD+SUSY event rates.}.

\section{Tagging procedure}

In this Section we  describe possible signatures of long-lived
gluinos in 4jet events at LEP I. We will resort to the fact
that gluinos should produce displaced vertices and offer a
complementary tool to the ALEPH study \cite{aleph}. 
We also single out  those 
combinations of SUSY parameters which could induce gluino decays into a 
large amount of missing energy. In this situation, the contribution of SUSY
events to the 4jet rate as selected in the ALEPH analysis would be
considerably reduced. 
Since we are implying that gluinos decay inside the LEP detectors, our
considerations will only apply to the case of lifetimes less than 
$10^{-9}$ sec or so. 
In terms of mass, we will focus our attention on the two regions:
(i) 0~${\buildrel{\scriptscriptstyle <}\over{\scriptscriptstyle\sim}}~
m_{\mathaccent"7E g}~{\buildrel{\scriptscriptstyle <}\over
{\scriptscriptstyle\sim}}$~1.5 GeV; 
(ii) $m_{\mathaccent"7E g}~{\buildrel{\scriptscriptstyle >}
\over{\scriptscriptstyle\sim}}$~3.5 GeV. 

Long-lived gluinos can be operationally defined as those which hadronise
before decaying. An inevitable consequence is that they live confined
into bound states, generically called $R$-hadrons \cite{window}.
The lightest of these would probably be the neutral, flavour singlet
$({\tilde g}g)$ and $({\tilde g}uds)$ hadrons.

If the gluino mass falls in the range (i),
decays into the minimum
hadronic mass, i.e., $(\tilde gg) \to \tilde\gamma \pi\pi$ and/or
$(\tilde g uds) \to \tilde\gamma \Lambda$, maximize the missing energy
and, therefore, the SUSY signal
in the ALEPH analysis can be well attenuated.
If $\tau_{\tilde g}\gg
\tau_b$, SUSY events should show displaced vertices at
a distance $d$ significantly larger than the decay length produced by 
a $b$-quark
(i.e., 300 $\mu$m or so). If $\tau_{\tilde g}\approx\tau_b$ then the 
`degeneracy'
discussed in Ref.~\cite{physlett} would occur between heavy quarks and 
gluinos, so that the double vertex tagging procedure combined with 
appropriate kinematic cuts (see Section 4) should help to disentangle
the gluino contribution. If $\tau_{\tilde g}\ll\tau_b$, then it may not be 
useful to look for detached vertices,
since these would not be detectable for $(\tilde g uds) \to
\tilde\gamma \Lambda$ and would be too close to the interaction
region. In general, for a gluino with mass below 1.5 GeV,
the fragmentation function should be similar to that for the charm quark, i.e.,
with $\langle z \rangle < 0.5$, or even softer if the mass is very
light.  Hence, there is a maximum missing energy possible, and it may
be that this whole region is excluded by the ALEPH analysis. If not,
there must be significant missing energy correlated with the
directions of the soft visible gluino jets. A combination of a
suitable missing energy distribution with the ALEPH analysis should be
able to find or exclude such a light gluino.

In the second regime (ii), the gluino would presumably fragment into a
$(\tilde gg)$ or $(\tilde g uds)$ hadron with a fragmentation function
perhaps similar to a $b$-hadron, i.e., with $\langle z \rangle \sim
0.75$. This hadron would then decay into a $\tilde \gamma$ plus
hadrons with a distribution similar to that for $\tilde g \to \tilde
\gamma q \bar q$. The missing energy would be maximized if the mass of
the $\tilde \gamma$ is close to that of the $\tilde g$, but a limit to
the mass difference is set by the requirements that the decay occurs
inside the tracking volume and that the squark masses be reasonable.
The charged multiplicity distribution should be similar to that for
$e^+e^- \to q \bar q$ at the same $q \bar q$ mass, and events occurring
inside $R \sim 0.5$~m with non-zero charged multiplicity should be
observed with high probability. Such events would have two hard jets,
two soft jets, missing energy, and two largely detached vertices with
$d\gg0.3$ mm
in the directions of the soft jets. If the rate corresponding to
these events is rather poor, then it is conceivable  
that the LEP collaborations might not have noticed them so far.
 
\section{Results}

\subsection{Production rates}

In this Section we compare the production rates of SUSY events, as a function
of the gluino mass in the range between 0 and 20 GeV,
to the yield of ordinary QCD events. We do this with and without assuming
vertex tagging
(the titles $e^+e^-\rightarrow$~VVjj and $e^+e^-\rightarrow$~jjjj in the 
forthcoming plots 
will refer to these two cases, respectively).
In the results of the cross sections for the untagged case  a
summation over all the quark flavours (massless
and massive) is implicit, whereas  for the tagged case 
we will consider the detached vertices as produced by gluinos and $b$-quarks
only, thus neglecting the case of $c$-decays\footnote{We assume the 
rate due to misidentification of gluons and massless quarks as heavy partons
negligible \cite{deangelis}.}, and sum over the remaining quark flavours.

In principle, one should also retain $c$-quark events among those
producing a detached vertex, eventually combining the corresponding rates
with those for $b$-quarks, according to the values of efficiency and purity
of the experimental analyses.
In fact, the lifetime of $c$-quarks is finite (around 1/3 of that of
the $b$'s) and is thus responsible for secondary vertices. Therefore, one
should expect that,
for values of the gluino lifetime around $1/3\tau_b$ (and below), charmed
hadrons can represent an additional background from ordinary QCD to the
SUSY signal and the actual positions
of the decay vertices of $c$- and $b$-quarks can partially overlap.
This is the reason why the algorithms determining the efficiency/purity of
flavour tagging used by the LEP~I collaborations contain a 
multiplicity selection rule (the number of tracks produced being 
higher for $b$'s than for $c$'s) \cite{revb}. 
The values of purity obtained in a single $b$-tag at LEP~I,
around $95\%$ or more \cite{deangelis,Squarcia,revb},
imply that above the $b$-selection cuts (in multiplicity and decay distance)
the ordinary QCD contribution is indeed almost entirely due to decaying
bottom hadrons, whereas below those cuts charmed hadrons  are mainly 
responsible for secondary vertices.
In other terms,
the $c$- and $b$-contributions would enter in our analysis `separately'
from each other into the QCD background to SUSY signals,
the relative small contamination being eventually established by the
experimental tagging strategy.
For reasons of space,
in the following we will illustrate the interplay between gluinos
and bottom quarks only. However, it must always be intended that in presence
of a short decay distance and/or a low secondary vertex multiplicity 
the actual
rates from ordinary QCD events will in the end need the inclusion of the
mentioned corrections due to the differences between $c$- and $b$-quarks.

In Fig.~4 we study the effect
of a non--zero gluino mass in the total cross section of the process
$e^+e^- \to qq\mathaccent"7E g\mathaccent"7E g$ for the  schemes
described in Section 2 and for different values of the $y_{\mathrm{cut}}$ 
parameter.
For $m_{\mathaccent"7E g} 
\buildrel{\scriptscriptstyle >}\over{\scriptscriptstyle\sim} 5$
 GeV the cross section falls exponentially
in the J and D schemes. For the G scheme the exponential behaviour
starts somewhat earlier, at 
$m_{\mathaccent"7E g}\buildrel{\scriptscriptstyle >}
\over{\scriptscriptstyle\sim} 2$ GeV,
and for the E scheme later, when
$m_{\mathaccent"7E g}\buildrel{\scriptscriptstyle
 >}\over{\scriptscriptstyle\sim}~7$ GeV. We already know that
SUSY rates certainly compare rather poorly to both the $q\bar qgg$
and $q\bar  q q'\bar q'$ contributions, if all quark flavours are 
retained and energy ordering is adopted \cite{masses}.
Nonetheless, one of the salient features in Fig.~4 is
that for $m_{\mathaccent"7E g}~{\buildrel{\scriptscriptstyle <}
\over{\scriptscriptstyle\sim}}~10$ GeV the mass suppression on the SUSY
rates is always less than one
order of magnitude. This is true independently of jet algorithm 
and for three typical values of the resolution
parameter $y_{\mathrm{cut}}$.
In any case the contribution for $m_{\mathaccent"7E g}>10$ GeV begins to be 
very small. Therefore, in the remainder of the paper we will confine ourselves 
to gluino masses up to 10 GeV only. At this point, one should recall that
the ordinary QCD production rates are much larger than the SUSY ones
displayed in Fig.~4. For example, 
when no vertex tagging is exploited and all flavours
are retained in the sample,
at the minimum of the 
$y_{\mathrm{cut}}$'s used there, one gets 
$\sigma(q\bar qgg)=4153(4498)[5862]\{9312\}$ pb and
$\sigma(q\bar qQ\bar Q)=187(217)[300]\{548\}$ pb, in correspondence
of the J(E)[D]\{G\} scheme. A similar pattern in the relative composition of 
4jet events persists also at larger values of the resolution
parameter. 

The cross sections as a function of $y_{\mathrm{cut}}$
for the different subprocesses yielding two displaced vertices
are presented in Fig.~5. The ratio between SUSY and pure
QCD events is clearly improved, so that $b\bar bq\bar q$
and gluino rates now compare to each other.
The largest contribution still comes from the subprocess $e^+e^- 
\to b\bar{b}gg$,
which is almost one order of magnitude larger than the other two
in the whole range of $y_{\mathrm{cut}}$. 
How to ameliorate this situation
will be discussed below. 
The gluino rates are shown
for three reference masses, $m_{\mathaccent"7E g} = 1,5, 10$ GeV. 
The pattern recognised in Fig.~4 as a function of the gluino mass is
also visible in Fig.~5 for the $y_{\mathrm{cut}}$ dependence. That is,
as the gluino mass increases the production rates
get smaller, however still remaining within the same order of magnitude if
$m_{\mathaccent"7E g}~
{\buildrel{\scriptscriptstyle <}\over{\scriptscriptstyle\sim}}~10$ GeV.
In practice, gluinos in the mass range up to 10 GeV have all sizeable 
production rates at LEP I in the jet schemes considered for usual values
of the jet resolution parameters.
This is indeed encouraging, as this means that the 4jet sample
could well be sensible to values of $m_{\mathaccent"7E g}$ larger than those
usually considered (i.e., of the order of the $b$-mass or below).

Fig.~6  shows the improvements that can be achieved with
heavy flavour tagging combined with the typical kinematic behaviour 
of gluino events, see Ref.~\cite{physlett}. 
For reference, the gluino mass has been fixed at 5 GeV,
though the main features of the plots do not depend on $m_{\mathaccent"7E g}$
as these are connected only to the fact that gluinos are always secondary 
products. Only the D scheme is shown, for the other schemes
exhibit very similar behaviour. The variable $Y_{ij}$ is the invariant scaled mass
\begin{equation}
Y_{ij}= \frac{(p_i+p_j)^2}{s}
\label{Yij},
\end{equation}
where $s$ is the center of mass energy ($s=M_Z^2$) and the indices
$ij$ label the jets as follows: (12) refer to the two vertex
tagged jets, and to the most energetic
jets in the case of energy ordering; (34) corresponds to the two
remaining jets. The distributions are normalised to one.
Note that in Fig.~6a
the $2\mathaccent"7E g 2q$  and $2b2q$ events
are peaked at low $Y_{12}$ while the $2b2g$ events are evenly distributed.
The peak in the first two cases is easily understood as it comes from
the propagator $g^*\to b\bar{b}/\mathaccent"7E g\mathaccent"7E g$,
 which is not present in
the third case (the tagged jets there come always from the $Z$ decay).
The long tail of the $2b2q$ spectra comes from the fact that there can be
`mis--tags' of $b$'s coming from the $Z$ propagator. The peak for
$2q2\mathaccent"7E g$ is even narrower, as the two gluinos are always produced
through gluon splitting, apart from a small contamination of
mis-tags coming from $2b2\mathaccent"7E g$. 
The strategy is now clear: for
$Y_{12}<0.2$ most of the SUSY signal is retained while $2b2q(2b2g)$ events
are reduced roughly by a factor of two(four).
In contrast, when energy ordering is performed (Fig.~6b) this effect
is washed out, as all the distributions have a similar shape and no useful
cut can be devised. Note that the distributions are finite due to
the masses of the tagged jets so that loop corrections will not change
significantly the behaviour presented here. The situation is even better 
if we look at Fig.~6c: the distribution for $2q2\mathaccent"7E g$ is flat 
and the other two distributions are peaked at $Y_{34}=0$. 
This effect is just the
complementary of Fig.~6a: the  (34) jets come from the $Z$ propagator
in the $2q2\mathaccent"7E g$ events while show the peak of the gluon splitting
for the other two cases. Again, when energy ordering is performed,
Fig.~6d,  the effect is wiped off.
Therefore, we adopt the following requirements to optimise the SUSY signal
over the ordinary QCD background: $Y_{12}<0.2$ and $Y_{34}>0.1$.
Note that the use of
the tagging procedure has been crucial for such an achievement. These simple
invariant mass distributions serve the purpose of reducing 
the ordinary QCD rates in case of 
$\tau_{\tilde g}~
{\buildrel{\scriptscriptstyle <}\over{\scriptscriptstyle\sim}}~\tau_b$,
as it can happen when 
$m_{\tilde g}~{\buildrel{\scriptscriptstyle <}\over{\scriptscriptstyle\sim}}
~1.5$ GeV. For $m_{\tilde g}~{\buildrel{\scriptscriptstyle >}
\over{\scriptscriptstyle\sim}}~3.5$ GeV,
where $\tau_{\tilde g}\gg\tau_b$ and the gluino and quark vertices are 
in principle well 
distinguishable, the kinematic distributions would clearly help to 
elucidate the underlying SUSY dynamics. 

In Fig.~7 we show the different contributions to the total
cross section in our tagging procedure like in Fig.~5, but
with the improved sample. The $2b2g$ event rates,
which were one order of magnitude larger than those of the other two
partonic components,
have been greatly reduced. All contributions are now comparable
(at least for $m_{\mathaccent"7E g}= 1\div5$ GeV). 
For $m_{\mathaccent"7E g}~
{\buildrel{\scriptscriptstyle >}\over{\scriptscriptstyle\sim}}~5$ GeV the 
ordinary QCD events can be most likely
eliminated
from the sample already on a displaced vertex basis, by asking, e.g.,
that the decay length is much longer than 0.3 mm. However, 
for completeness we 
report the rates for large gluino masses too, as the tagging procedure
could be complicated by the fact that a large part of the vertex tagged 
hadronic sample at LEP I has been collected via a bi-dimensional 
tagging \cite{bidimensional}. Therefore, projections of
different decay lengths $d$ could well appear the same on the
reproduced event plane. It is also worth recalling that the 
fact that gluinos are electrically neutral whereas quarks are charged can 
hardly
be useful in 4jet analyses as there is extremely low efficiency in
measuring the total jet charge, especially in multijet events. That explains 
why, for instance, this difference is not  used to discriminate partonic 
compositions in ordinary 4jet events (as gluons too are neutral). 
In summary, we have shown that it is feasible to significantly
enhance the signal of possible light gluino species over the
QCD background using tagged samples with the help of elementary kinematical
distributions.

\subsection{Missing energy distributions}

In this section we study the decays rates of SUSY events, for
	three representative values of the gluino mass which yield sizeable
production rates. In particular, we will investigate the spectrum in missing
energy inside the gluino jets, trying to establish the quantitative relevance 
in the total SUSY sample of hadronic events carrying an energetic imbalance
that does not meet the usual trigger thresholds of the LEP I detectors.

In discussing the possible decay modes of the gluino, two assumptions 
need to be made. The first is the condition of $R$-parity conservation. The 
second is the choice of the lowest mass Supersymmetric particle.
$R$-parity, defined to be even for ordinary particles and odd for their 
Supersymmetric counterparts, needs to be preserved if lepton and baryon 
numbers are exactly conserved. This implies that the lightest 
Supersymmetric particle is exactly stable. 
In this paper we shall take it for granted that the neutralino (photino)
is the lowest mass Supersymmetric particle. Failing this condition, the
next likely choice would be the case where the scalar neutrinos are lower in
mass. However, very light doublet sneutrinos are excluded by the $Z$ 
width constraints \cite{PDG}.

The choice of the scalar quark masses ${\mathaccent"7E M}_L$ and 
${\mathaccent"7E M}_R$ \cite{Nilles} 
affects the gluino branching ratios. In 
particular, considering only one flavour of massless quarks and assuming
that the photino is massless, the ratio between the widths of the
two dominant gluino decay modes is given by:
\begin{equation}\label{BR}
\frac{\Gamma({\tilde{g}\rightarrow g\tilde{\gamma}})}
     {\Gamma({\tilde{g}\rightarrow q\bar{q}\tilde{\gamma}})}=
\frac{3\alpha_s}{4\pi}
\frac{({\mathaccent"7E M}_R^2-{\mathaccent"7E M}_L^2)^2}
{({\mathaccent"7E M}_L^4+{\mathaccent"7E M}_R^4)}.
\end{equation}
Therefore, the quark-antiquark-neutralino decay channel
is  dominant over the gluon-neutralino one. However, 
in some SUSY models
the $L$ and $R$ mass eigenstates may differ by a factor of two or even 
more, such that ${({\mathaccent"7E M}_R^2-{\mathaccent"7E M}_L^2)^2}/
{({\mathaccent"7E M}_L^4+{\mathaccent"7E M}_R^4)}\buildrel
{\scriptscriptstyle >}\over{\scriptscriptstyle\sim}1/2$ \cite{Haber-Kane}.
Furthermore, as the photino mass approaches that of the gluino,
$m_{\mathaccent"7E \gamma}/m_{\mathaccent"7E g}\rightarrow1$, 
the three-body decay 
mode suffers 
a further suppression, which goes as 
$(1-m_{\mathaccent"7E \gamma}/m_{\mathaccent"7E g})^2$. 
A more extensive review of the gluino decay channels can be found, 
for example, in Section 
3.4 of Haber and Kane \cite{Haber-Kane} (see also references therein).

We have computed the relevant decay currents by using
helicity amplitude techniques, and incorporated these into a complete
matrix element for gluino production and decay, over the 
appropriate phase space. In doing so,
two different formalisms were employed: the usual helicity 
projector method \cite{IZ} and the techniques of Ref.~\cite{KS}. The results
obtained with the two methods agree for any polarisation state
if in the latter formalism
one modifies the helicity projections to coincide with 
the physical choice along the direction of the final partons.

Before studying the decay spectra, a few comments are in order concerning
the fragmentation of a gluino. As already mentioned, a gluino would appear 
at the end of a hadronisation process confined into a 
bound state. 
Now, the decay kinematics of $R$-hadrons is 
in principle different from that of free ${\mathaccent"7E g}$'s.
However, if the gluino is sufficiently heavy (say, $m_{\mathaccent"7E g}~
{\buildrel{\scriptscriptstyle >}\over{\scriptscriptstyle\sim}}~3.5$ GeV 
\cite{Haber-Kane}),
the phenomenology of the decay products of such $R$-hadrons 
would be similar to that of unbounded gluinos. 
In particular, the basic result is that the 
$\mathaccent"7E  \gamma$ energy spectrum
roughly agrees with that produced by a freely decaying $\mathaccent"7E g$
as long as 
$m_{\mathaccent"7E \gamma}/m_{\mathaccent"7E  g}$ is not too close to one 
\cite{ACCMM}.
For lighter gluinos the analysis is less straightforward. However, 
according to Ref.~\cite{Franco}, it is reasonable
to expect that these SUSY hadrons would again decay similarly to free gluinos, 
provided that $m_{\mathaccent"7E g}$ is replaced 
by an `effective' $R$-hadron mass 
equal to $\approx 0.75~m_{\mathaccent"7E g}$.
For our purposes, we assume that the mass appearing in the decay spectra
is in fact that of the SUSY parton in the mass range (ii), whereas in the
interval (i) it represents the mentioned effective mass. 
Furthermore, on the one
hand, we confine ourselves to values of $m_{\mathaccent"7E \gamma}$ strictly
smaller than $m_{\mathaccent"7E g}$ in order to maintain valid our 
approximation over the range 
$m_{\mathaccent"7E g}~
{\buildrel{\scriptscriptstyle <}\over{\scriptscriptstyle\sim}}~1.5$ GeV;
on the other  hand, we will push the ratio 
$m_{\mathaccent"7E \gamma}/m_{\mathaccent"7E g}$ up to 3/4 in order
to maximise the amount of missing energy carried away by the undetected
photino.

The results we have obtained for the energy distribution of the missing
energy after the two decays are displayed in Figs.~8a--c and
Figs.~9a--c (in correspondence of the two possible decays).
The crucial point is that the amount of missing energies
produced could be so large that SUSY events of the type
$q\bar q {\tilde g}{\tilde g}$ are not recognised as 4jet events.
In fact, experimental analyses have a minimal hadronic energy cut on
each of the four jets, in order to reduce the
background due to poorly reconstructed events.

The $E_{\mathrm{miss}}$ spectra 
are shown for four kinematical decay configurations:
a massless photino, and a massive one 
with $m_{\mathaccent"7E  \gamma}
=n/4m_{\mathaccent"7E g}$, with $n=1,2,3$, 
and for three gluino masses $m_{\mathaccent"7E g} =1,5,10$ GeV.
It is clear from both Fig.~8a--c and 9a--c that the missing energy
spectrum gets harder as $m_{\tilde g}$ and $m_{\tilde\gamma}$ increase
in both decay channels considered. The effect is common to all
algorithms. 
For $m_{\tilde g}=1$ GeV the mean value of the missing energy is
always below 10 GeV in both decay channels, and it can grow up
to more than 15 GeV if $m_{\tilde g}=10$ GeV and $m_{\tilde \gamma}=7.5$
GeV. Under such conditions, it could be argued
that $q\bar q {\tilde g}{\tilde g}$ events can pass unobserved
as actual 4jet events if tight constraints are implemented 
on the missing mass energy of the hadronic event sample.

\section{Summary and conclusions}

In this paper we have studied the production and decay rates
of $e^+e^-\to q\bar q{\tilde g}{\tilde g}$ events at LEP~I, where
${\tilde g}$ represents a relatively light (up to 10 GeV in mass)
and long-lived (up to $10^{-9}$ sec in lifetime) gluino,
and compared these to the yield of ordinary QCD events of the type
$e^+e^-\to q\bar qgg$ and $e^+e^-\to q\bar qq'\bar q'$, involving quarks
$q$ and gluons $g$. The presence of such SUSY events in 4jet samples at LEP~I
has been advocated in the past years to explain the disagreement between
the values of the strong coupling constant $\alpha_s$ as measured
from the deep-inelastic scattering and the $Z$-peak data.  This was further
motivated by the initial discrepancy between the QCD predictions
for the colour factors $C_A$, $C_F$ and $T_F$ and their actual measurements
obtained in earlier analyses \cite{colour} by the LEP collaborations, as these
constants are sensitive to additional SUSY contributions. The
claim about the possible existence of gluinos in LEP I data has apparently
become less
convincing during the recent two or three years, 
as the experimental and theoretical analyses of the data have reached
a higher level of sophistication and precision. 
Very recent studies seem to exclude 
gluinos with masses up to 6.3 GeV.
Although such results represent clear progress towards settling the
ongoing dispute about the existence of SUSY signals at LEP I, we have outlined 
here a complementary approach guided by two considerations.

First, in all the mentioned analyses no vertex tagging was exploited
in assigning the momenta of the jets to the corresponding partons
from which the former originate. The study of 4jet events showing two 
secondary vertices produced in the decay of $c$- and $b$-quarks has
in fact been
proved to be successful in reducing the error on the QCD
colour factor which is most sensitive to the possible presence of light
gluinos (that is, $T_R=N_FT_F$). Furthermore, we have  
also shown that simple kinematic distributions (such as the invariant
masses of the two vertex tagged jets and of the remaining two) can 
effectively 
help to enrich significantly the 4jet samples of gluino events (if existing). 
In fact, the latter, on the one hand, should yield displaced vertices and, on 
the other hand, are always produced as secondary partons (contrary to heavy 
quarks). 

Second, the validity of the 
result quoted by the ALEPH collaboration (the most constraining one)
could be undermined by the fact that, in their procedure of
selecting candidate 4jet samples, events carrying 
a large fraction of missing energy were not included. As a matter 
of fact, gluinos (or better, $R$-hadrons, in which the SUSY partner of the
gluon is confined) should predominantly decay into `photinos', which escape
detection.  Indeed, there are kinematic configurations in which the ratio
between the gluino and photino mass is such that the missing energy is
rather large and, conversely, the left-over one for the hadronic system
arising from the SUSY decay is rather small, such that these events might
not pass the experimental 4jet resolution and selection criteria. 

In the above context, we believe to have obtained interesting results
for future studies. In fact, we have shown that, in the vertex tagged
sample of 4jets, SUSY events become comparable to the rates of ordinary
QCD events for gluino masses up to about 10 GeV, thus well beyond the
bounds presently set on this quantity. Furthermore, we have indicated that
the latter cannot be reliable if the photino mass is not negligible
compared to that of the gluino. Therefore, we conclude that experimental
analyses based on our approach should help in clarifying the present debate,
either contradicting the present bounds on the gluino mass 
or improving these by extending the experimental coverage of the
so-called `light gluino window'. 
For example, over the mass region $m_b
< m_{\tilde g}~
\buildrel{\scriptscriptstyle <}\over{\scriptscriptstyle\sim}~10$ GeV, SUSY
rates should still be sizable and yield an unmistakable signature
with two hard jets, two soft ones, large missing energy, two detached 
vertices with 
$c\tau~\buildrel{\scriptscriptstyle >}\over{\scriptscriptstyle\sim}~0.3$
mm, a neat peak in the invariant mass of
the vertex tagged dijet system and a very flat distribution in the mass
of the other two jets. 

In carrying out our study we have resorted to  parton level 
calculations, which include all masses of primary and secondary partons
exactly. Although
we have not implemented a full Monte Carlo procedure including
the fragmentation of the gluinos into hadrons or the decay
of the latter into jets and missing particles, we have used 
analytic approximations which should mimic well the actual gluino 
phenomenology to a degree of 
accuracy compatible with that of the current experimental analyses. In this 
respect, we have indicated possible signatures of gluinos decaying inside the 
LEP detectors, as a function of both the mass and the lifetime of the
SUSY particle.

Before closing, we would like to point out a few crucial aspects of our work.
First, contrary to many previous studies (which did not exploit vertex tagging
and/or kinematical cuts) in which the gluino component represents an effect
of just a few percent (thereby being of the same order as next-to-leading
and/or hadronisation corrections), we have been concerned 
with SUSY rates that are always comparable or even larger than those 
produced by pure QCD events. Therefore, the inclusion of the mentioned 
corrections will not spoil our results. Second, for values
of gluino masses up to 10 GeV or so,
our conclusions are essentially the same {independently} 
of the jet algorithm and of the value used for $y_{\mathrm{cut}}$ 
(although the actual cross sections and the behaviour of the distributions do
certainly depend on them). In the end, the magnitude of higher order
and hadronisation effects as well as 
experimental considerations will determine which algorithm and which
resolution parameter are the most suitable to use, though
the Geneva
algorithm seems to be slightly favoured due to its special sensitivity
to the actual number of
active flavours and a smaller scale dependence in NLO corrections.
Third, by adopting the current LEP I values of vertex tagging efficiency 
and luminosity, we should expect a statistically significant analysis,
based on several thousands of doubly tagged 4jet events.
Fourth, since those presented here are theoretical results from parton level
calculations, they will necessarily
have to be folded with detailed experimental 
simulations (including both fragmentation/hadronization and detector effects),
such that one could even improve at that stage our 
procedure: for example, by exploiting various differences (in charge, mass, 
lifetime) that occur between heavy quarks and gluinos.

We finally remark that in the long term our arguments could well be
of interest also to the SLC experiment at SLAC, as microvertex devices
are installed there and they are known to have achieved by now a considerable
tagging efficiency, so to hopefully compensate for the present lack of  
statistics of their data with respect to the LEP ones.

\section*{Acknowledgements}

We thank Bas Tausk and Val Gibson 
for valuable discussions. We are also grateful to Ben Bullock for carefully
reading the manuscript.
This work is supported in part by the
Ministero dell' Universit\`a e della Ricerca Scientifica, the UK PPARC,
the Spanish CICYT project AEN 94-0936, and  the EC Programme
``Human Capital and Mobility'', Network ``Physics at High Energy
Colliders'', contracts CHRX-CT93-0357 DG 12 COMA (SM) and 
ERBCHBICHT (RMT). 
KO is grateful to Trinity College and 
the Committee of Vice-Chancellors and Principals
of the Universities of the United Kingdom for financial support.

\goodbreak

\vfill
\newpage

\section*{Figure Captions}

\begin{itemize}

\item[{[1]}] Excluded regions of gluino masses and lifetimes (shaded areas), 
from Ref.~\cite{Kileng-Osland1}.

\item[{[2]}] Feynman diagrams at lowest order contributing to 
4jet production in $e^+e^-$ annihilations in ordinary QCD:
$2q2g$ contribution (a,b and c); $4q$
contribution (d). All possible permutations are not shown.

\item[{[3]}] Additional Feynman diagrams at lowest order contributing to 
4jet production in $e^+e^-$ annihilations in QCD+SUSY: $2q2\mathaccent"7E g$ 
contribution. The other permutation is not shown. 

\item[{[4]}] Cross sections for $2q2\mathaccent"7E g$ production.
A sum over the quark flavours $q=u,d,s,c$ and $b$ is implied. 
Curves are given as function of the gluino mass,
for the four different algorithms J, E, D and G introduced
in the text and three choices of 
the jet-scheme resolution parameter. 
The combinations are: $y_{\mathrm{cut}}^J=0.01(0.02)[0.04]$,
                      $y_{\mathrm{cut}}^E=0.01(0.02)[0.04]$,
                      $y_{\mathrm{cut}}^D=0.002(0.004)[0.008]$,
                      $y_{\mathrm{cut}}^G=0.02(0.04)[0.08]$, in   
                      continuous(dashed)[dotted] lines.

\item[{[5]}] Cross sections for the three different contributions
to 4jet production in QCD+SUSY, in case of vertex tagging.
Quark flavours are intended as summed over $q=u,d,s,c$ and $b$. 
Curves are given as function of the jet-scheme resolution parameter,
for the four different algorithms J, E, D and G introduced
in the text and three different values of gluino mass: $m_{\mathaccent"7E g}
=1(5)[10]$ GeV, in
dotted(chain-dashed)[chain-dotted] lines.

\item[{[6]}] (a) Differential distributions in the rescaled invariant mass 
of the vertex tagged pair of jets 
for the three different contributions
to 4jet production in QCD+SUSY.
Quark flavours are intended as summed over $q=u,d,s,c$ and $b$. 
Curves are given 
for the  algorithm D  introduced
in the text, for the minimum of the jet-scheme resolution parameter
$y_{\mathrm{cut}}=0.002$.
The gluino mass is set equal to 5 GeV. Distributions are normalised to unity.
(b) The same distributions in case of energy ordering of the jets. Here,
the invariant mass corresponds to that of the two most energetic jets.
An additional sum over $q'=u,d,s,c$ and $b$ is implied.
(c) Same differential distributions as (a) in the rescaled invariant mass 
of the `untagged' pair of jets.
(d) Same distributions as in (b) in the  rescaled invariant mass 
of the `untagged' pair of jets.

\item[{[7]}] Cross sections for the three different contributions
to 4jet production in QCD+SUSY, in case of vertex tagging, 
after the kinematical cuts.
Quark flavours are intended as summed over $q=u,d,s,c$ and $b$. 
Curves are given as function of the jet-scheme resolution parameter,
for the four different algorithms J, E, D and G introduced
in the text and three different values of gluino mass: $m_{\mathaccent"7E g}
=1(5)[10]$ GeV, in
dotted(chain-dashed)[chain-dotted] lines.

\item[{[8]}] Differential distributions in the missing
energy of the `gluino jet' after the SUSY decays  
$\mathaccent"7E g\rightarrow q\bar q\mathaccent"7E{\gamma}$, 
where $\mathaccent"7E{\gamma}$ represents the `photino'.
The masses of this latter are $m_{\mathaccent"7E{\gamma}}=0$ (continuous), 
$1/4m_{\mathaccent"7E g}$ (dashed), $1/2m_{\mathaccent"7E g}$ (dotted) and  
$3/4m_{\mathaccent"7E g}$ (chain-dotted).
The gluino masses are: $m_{\mathaccent"7E g}=1$ (a), 5 (b) and 10 (c) GeV.
The mass refers to the `effective' mass of the bound gluino if
$m_{\mathaccent"7E g}=1$ GeV. No kinematical cut is here applied.

\item[{[9]}] Differential distributions in the missing
energy of the `gluino jet' after the SUSY decays  
$\mathaccent"7E g\rightarrow g\mathaccent"7E{\gamma}$, 
where $\mathaccent"7E{\gamma}$ represents the `photino'.
The masses of this latter are $m_{\mathaccent"7E{\gamma}}=0$ (continuous), 
$1/4m_{\mathaccent"7E g}$ (dashed), $1/2m_{\mathaccent"7E g}$ (dotted) and  
$3/4m_{\mathaccent"7E g}$ (chain-dotted).
The gluino masses are: $m_{\mathaccent"7E g}=1$ (a), 5 (b) and 10 (c) GeV.
The mass refers to the `effective' mass of the bound gluino if
$m_{\mathaccent"7E g}=1$ GeV. 
Normalisations are to unity. No kinematical cut is here applied.

\end{itemize}
\vfill
\clearpage
\begin{figure}[p]
\centerline{
\epsfig{figure=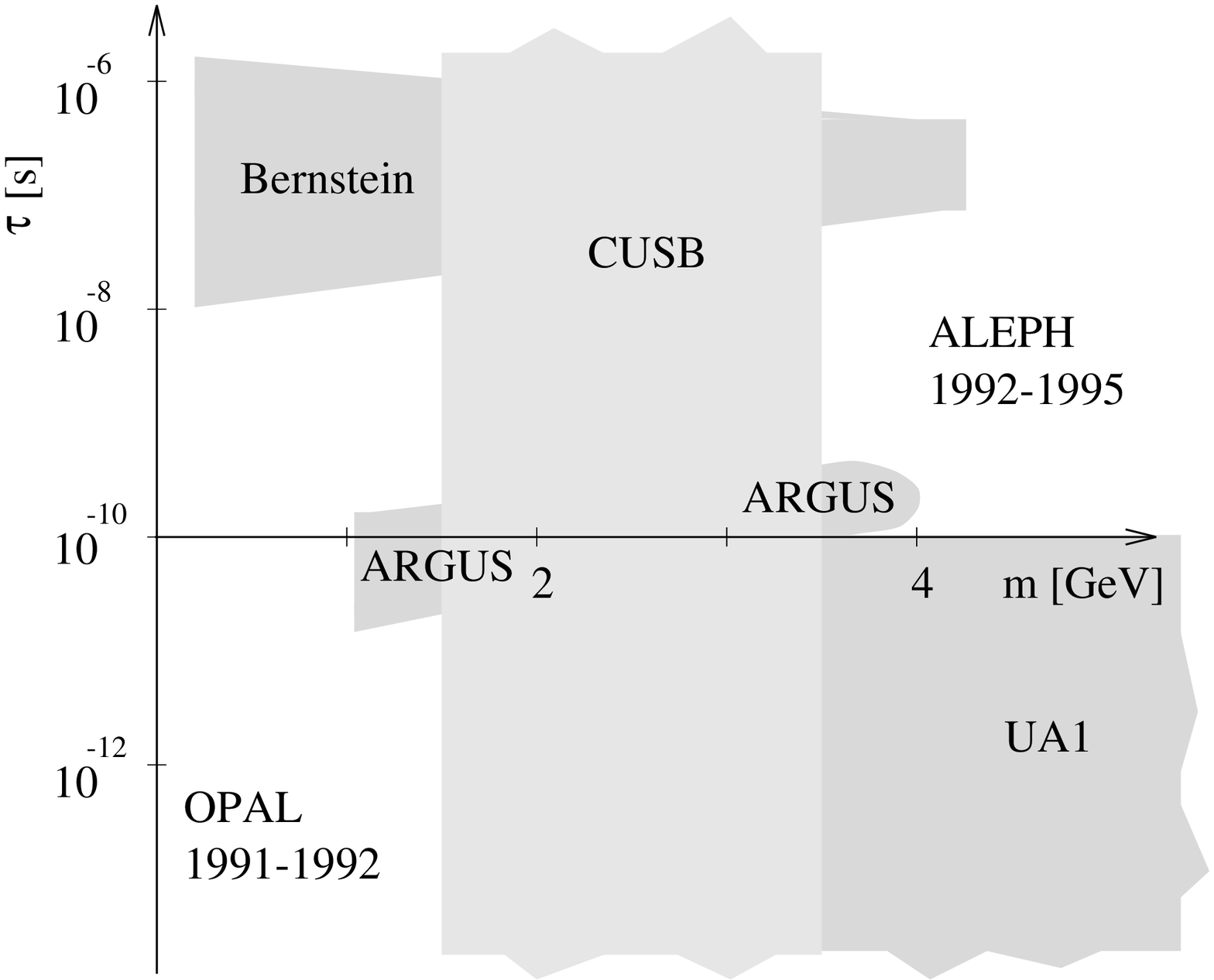,width=10cm}
}
\vspace*{2cm}
{\centerline{\Large{Fig. 1}}}
\end{figure}
\vfill
\clearpage

\begin{figure}[p]
~\epsfig{file=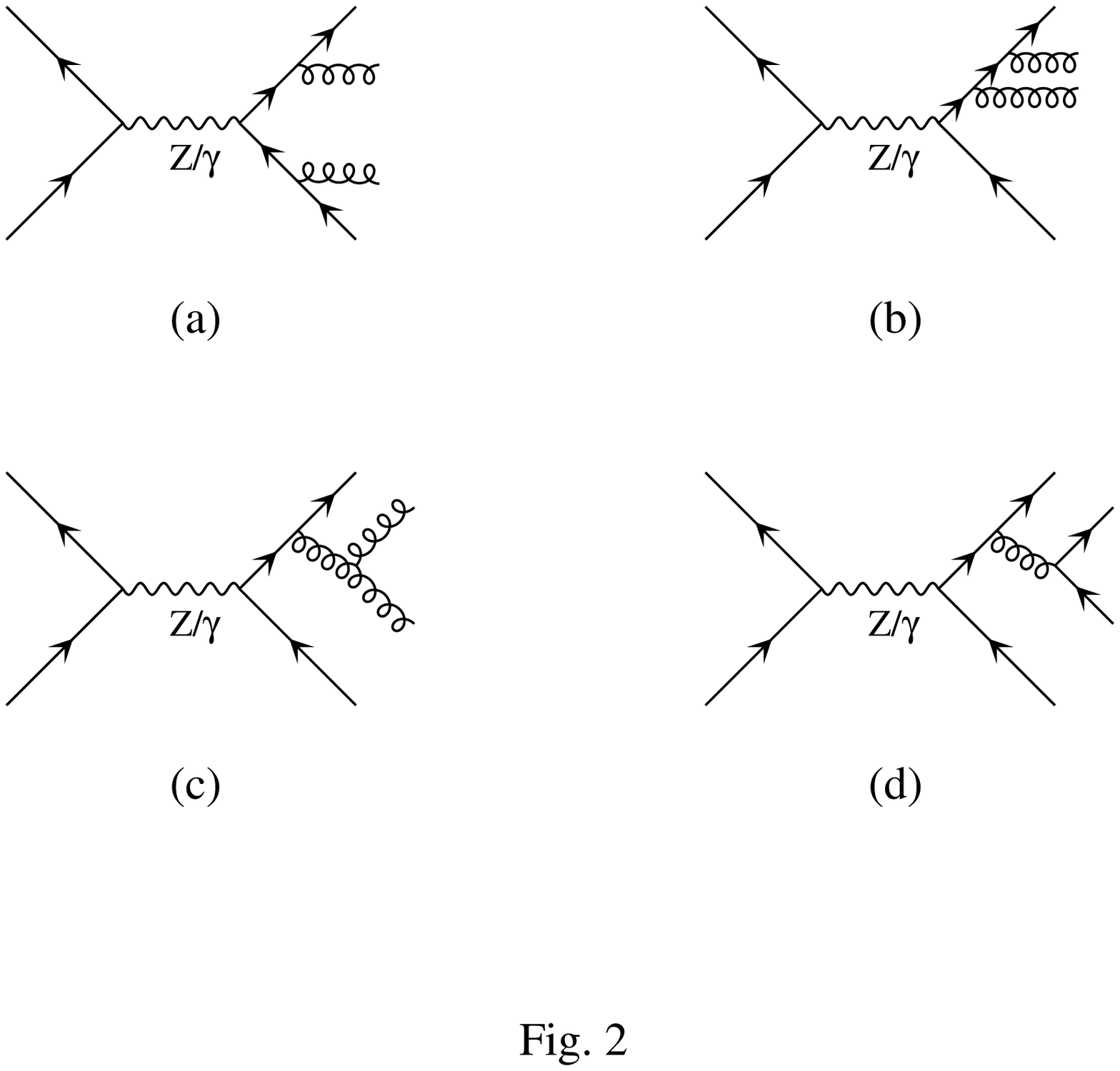,height=22cm}
\end{figure}
\stepcounter{figure}
\vfill
\clearpage

\begin{figure}[p]
~\epsfig{file=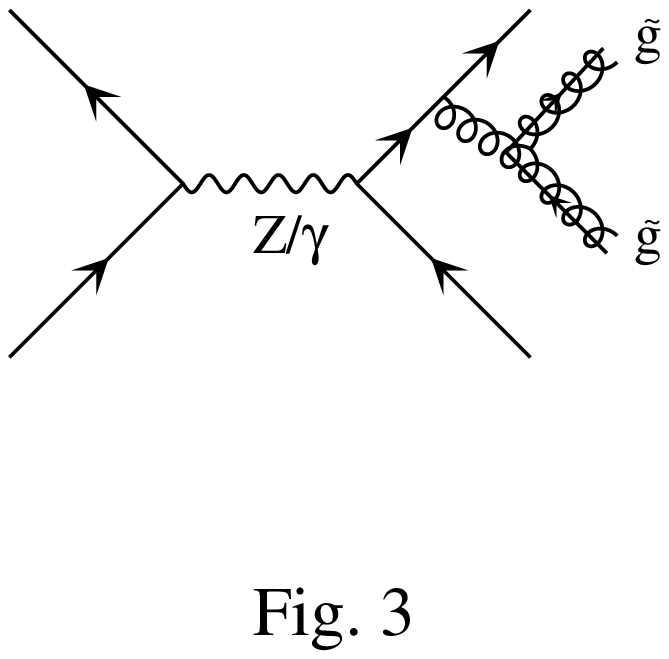,height=22cm}
\end{figure}
\stepcounter{figure}
\vfill
\clearpage

\begin{figure}[p]
~\epsfig{file=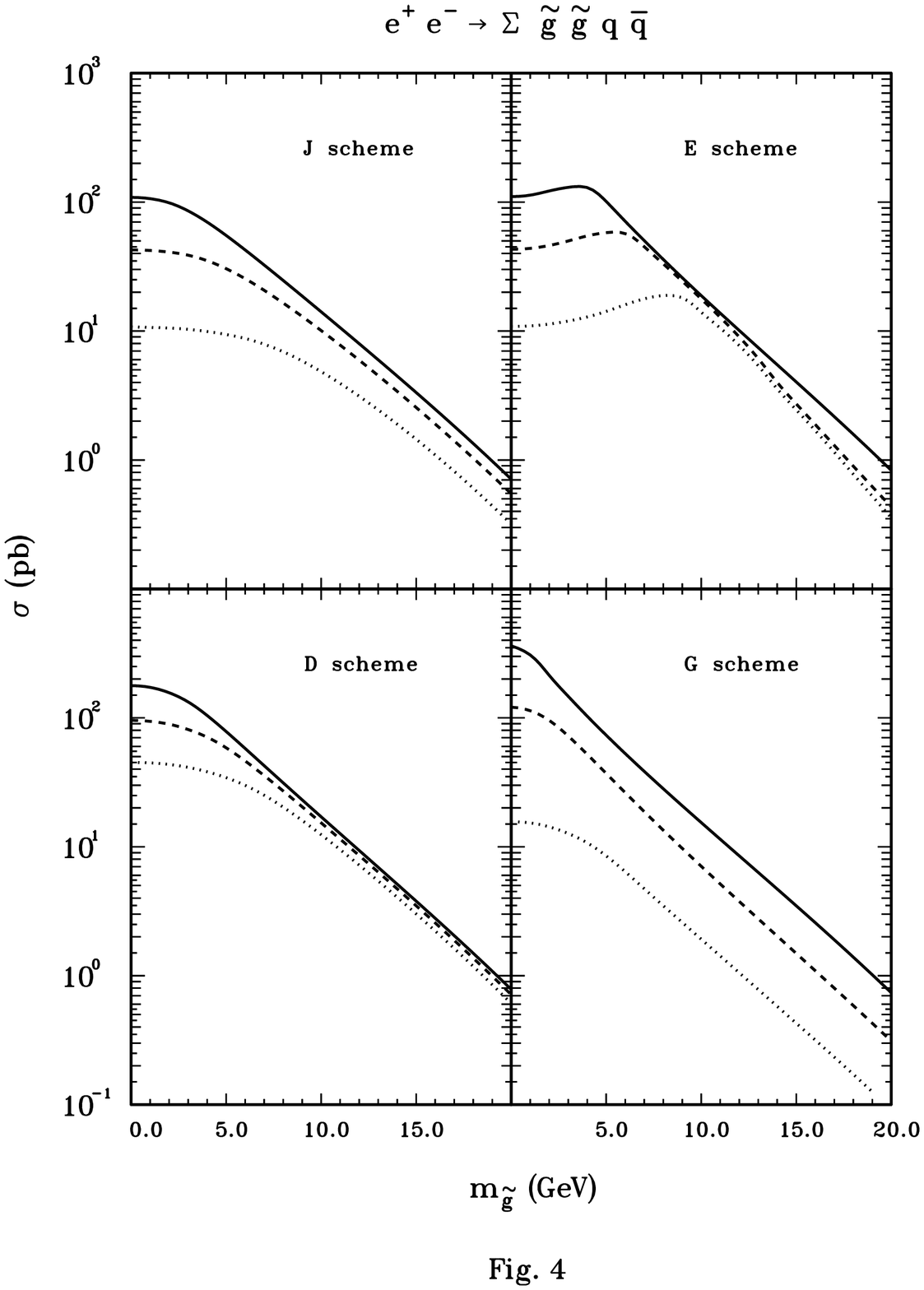,height=22cm}
\end{figure}
\stepcounter{figure}
\vfill
\clearpage

\begin{figure}[p]
~\epsfig{file=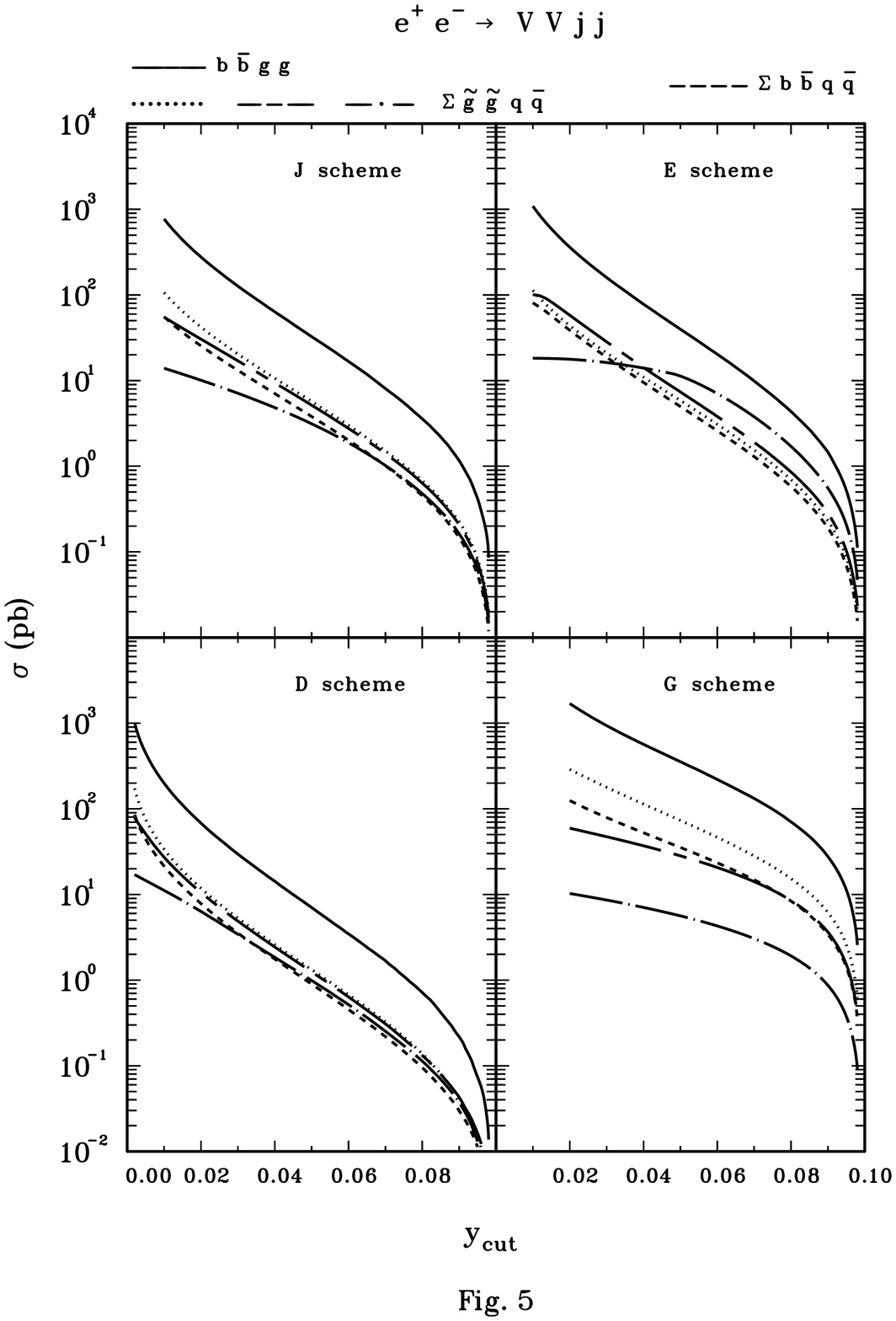,height=22cm}
\end{figure}
\stepcounter{figure}
\vfill
\clearpage

\begin{figure}[p]
~\epsfig{file=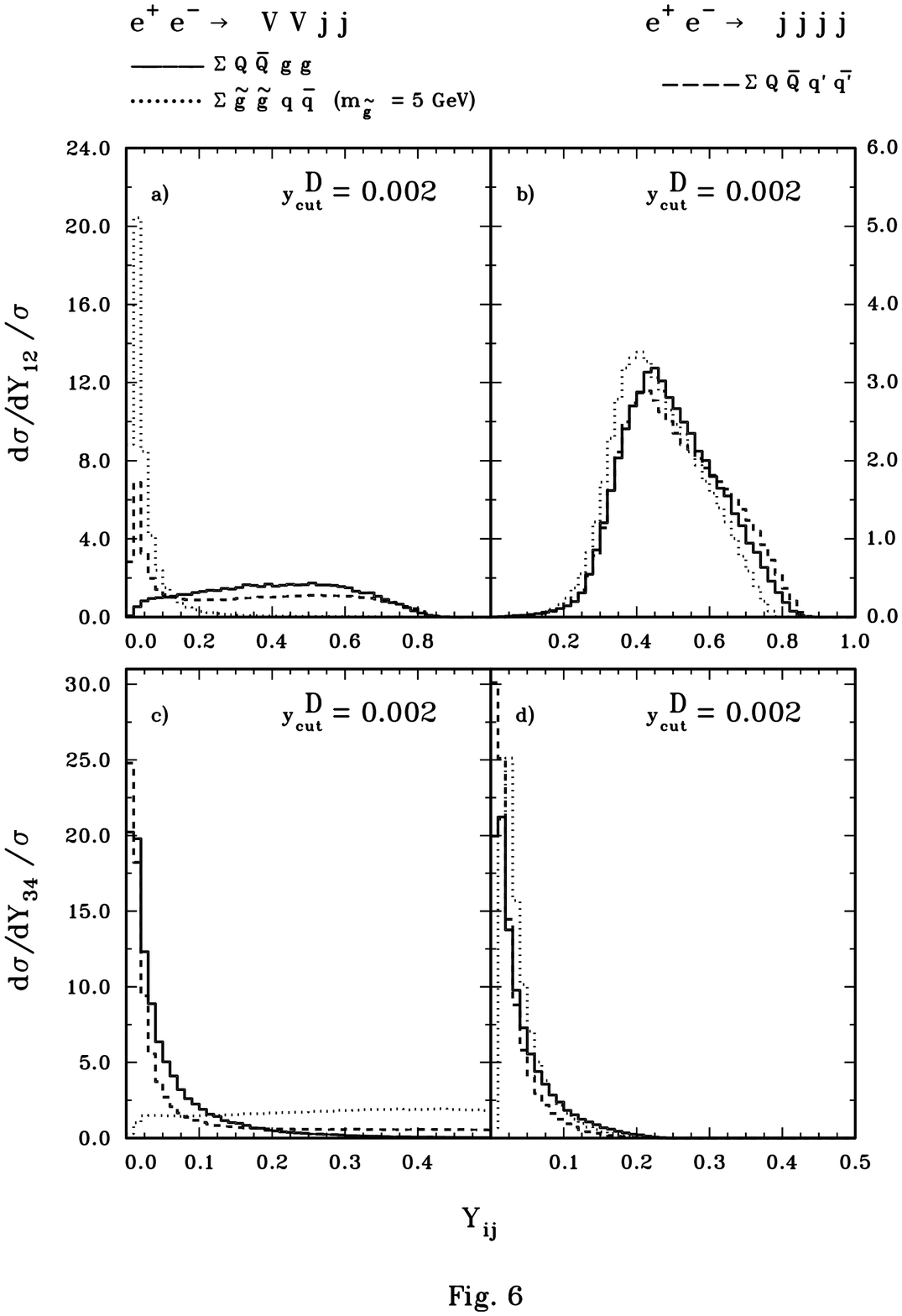,height=22cm}
\end{figure}
\stepcounter{figure}
\vfill
\clearpage

\begin{figure}[p]
~\epsfig{file=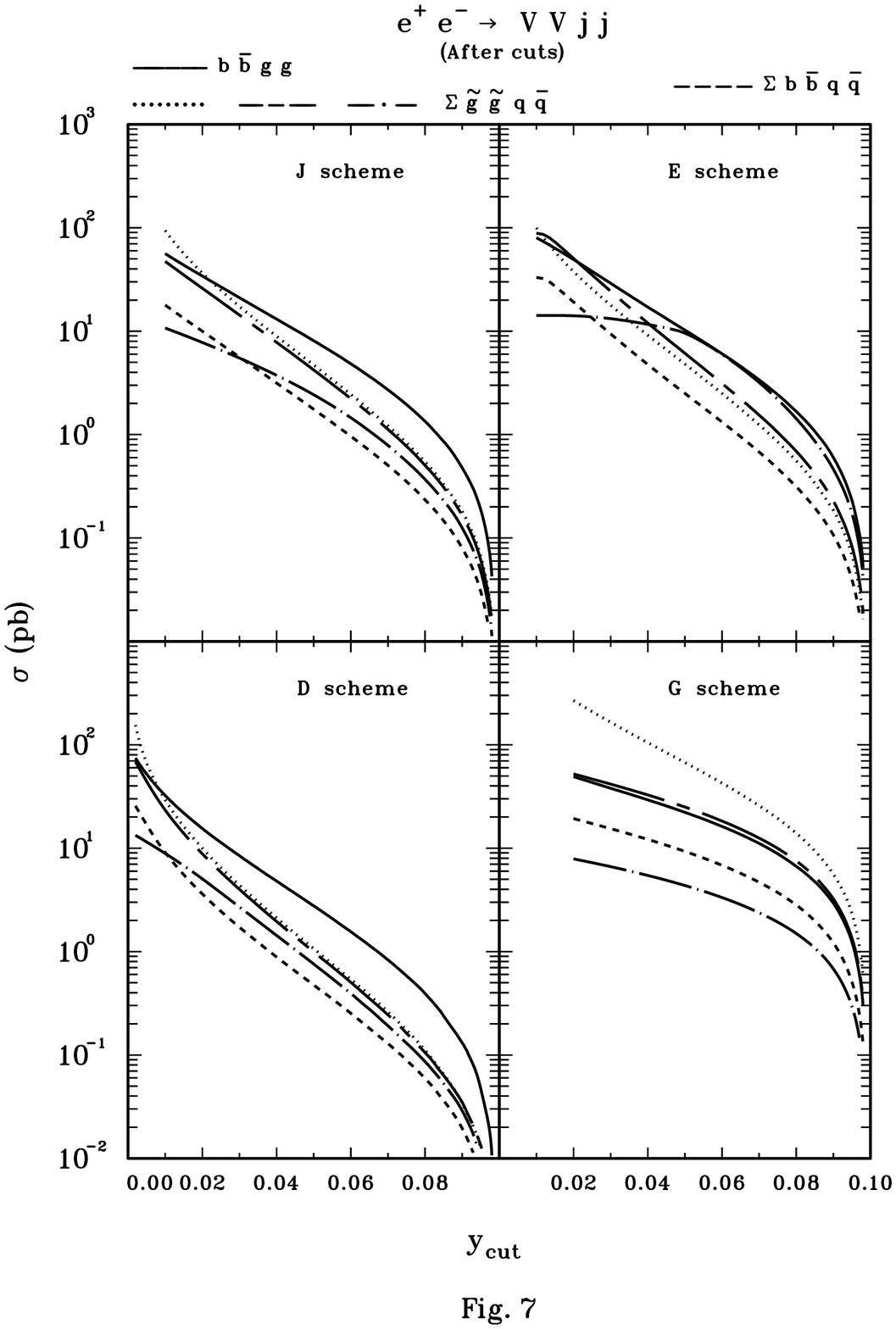,height=22cm}
\end{figure}
\stepcounter{figure}
\vfill
\clearpage

\begin{figure}[p]
~\epsfig{file=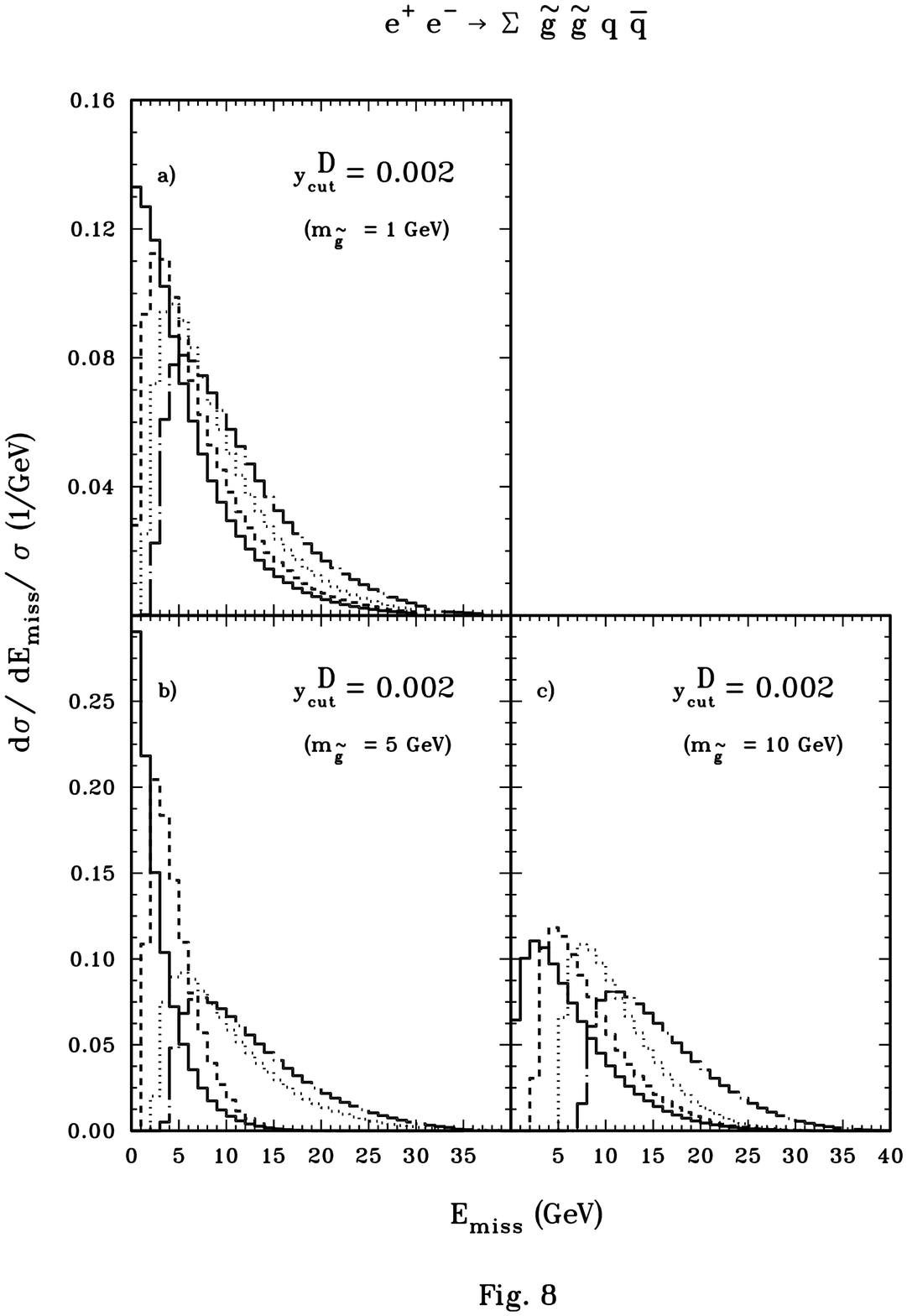,height=22cm}
\end{figure}
\stepcounter{figure}
\vfill
\clearpage

\begin{figure}[p]
~\epsfig{file=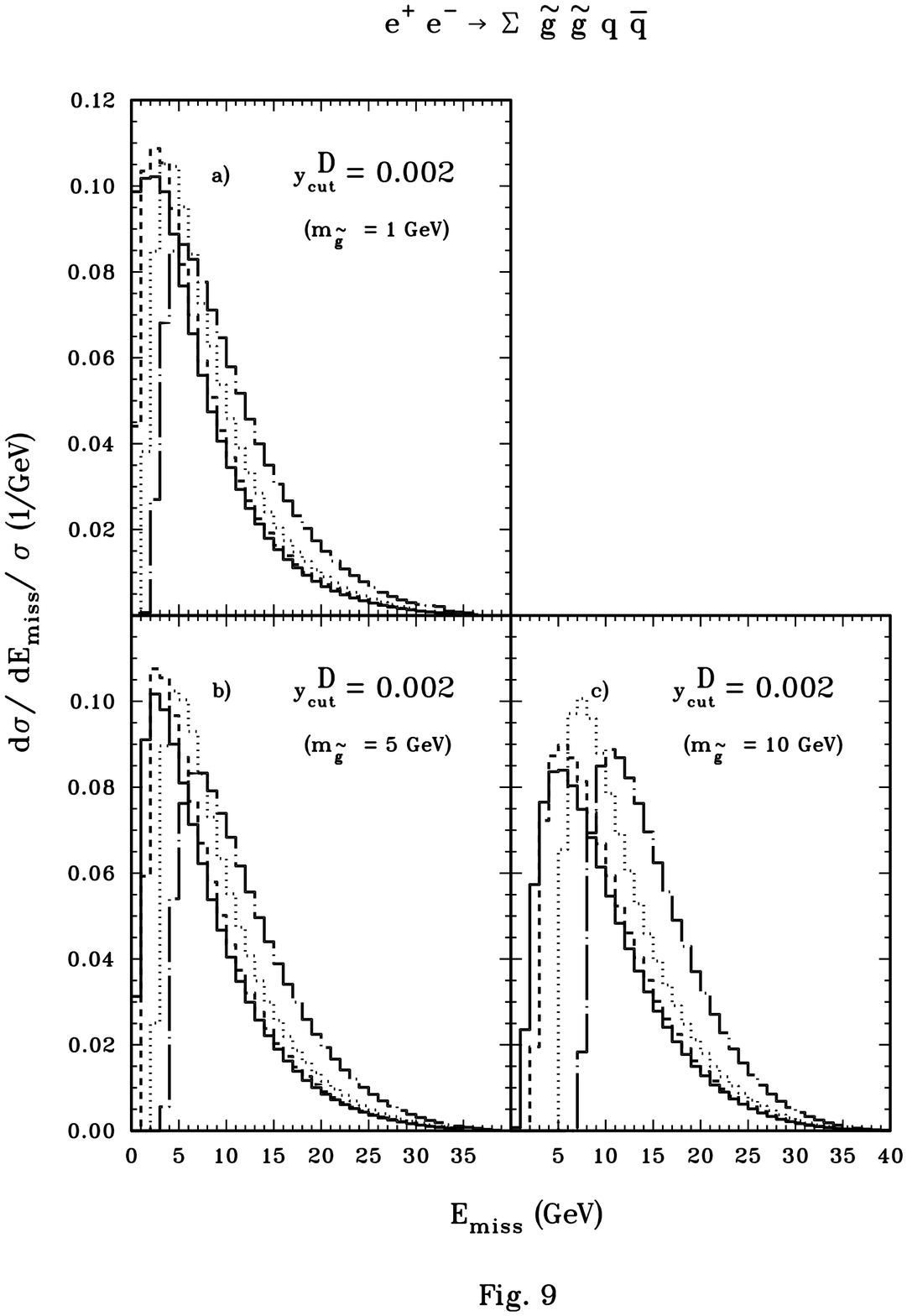,height=22cm}
\end{figure}
\stepcounter{figure}
\vfill

\end{document}